\title{Multi-loop atomic Sagnac interferometry}

\author{Christian Schubert$^{a,1,2}$, Sven Abend$^{2}$, Matthias Gersemann$^{2}$, Martina Gebbe$^{3}$,\\
Dennis Schlippert$^{2}$, Peter Berg$^{2}$, Ernst~M. Rasel$^{2}$\\  \\
\footnotesize $^{a}$ Christian.Schubert{\at}dlr.de\\
\footnotesize $^{1}$ Deutsches Zentrum für Luft- und Raumfahrt e.V. (DLR), Institut für Satellitengeodäsie und\\ 
\footnotesize Inertialsensorik, c/o Leibniz Universität Hannover, DLR-SI, Callinstraße 36, 30167 Hannover, Germany\\
\footnotesize $^{2}$ Institut für Quantenoptik, Gottfried Wilhelm Leibniz Universität Hannover, Welfengarten 1,\\
\footnotesize D-30167 Hannover, Germany\\
\footnotesize $^{3}$ Zentrum für angewandte Raumfahrttechnologie und Mikrogravitation (ZARM),\\
\footnotesize Universität Bremen, Am Fallturm, D-28359 Bremen, Germany \date{}
}

\newcommand{\abstractText}{\noindent
The sensitivity of light and matter-wave interferometers to rotations is based on the Sagnac effect and increases with the area enclosed by the interferometer.
In the case of light, the latter can be enlarged by forming multiple fibre loops, whereas the equivalent for matter-wave interferometers remains an experimental challenge.
We present a concept for a multi-loop atom interferometer with a scalable area formed by light pulses.
Our method will offer sensitivities as high as $2\cdot10^{-11}$\,rad/s at 1\,s in combination with the respective long-term stability as required for Earth rotation monitoring.\\ \\
\textbf{Keywords:} Atom interferometer, Matter-wave interferometry, Rotation measurement \\ \\
}

\documentclass[11pt, a4paper, twocolumn]{article}
\newcommand{\at}{\makeatletter @\makeatother}

\usepackage{sansmath}
\sansmath

\usepackage{xcolor}
\usepackage{titlesec}
\titleformat*{\section}{\bfseries}
\usepackage[
left=2.00cm,
right=2.00cm,
top=2.00cm,
bottom=2.00cm
]{geometry}
\usepackage[font=footnotesize,labelfont=bf]{caption}
\usepackage{datetime}
\RequirePackage{graphicx}
\usepackage{amssymb}
\usepackage{booktabs}
\usepackage{xurl}
\usepackage[numbers,comma,sort&compress]{natbib}
\usepackage{abstract}

\usepackage{hyperref}
\hypersetup{colorlinks=true, urlcolor=blue, linkcolor=blue, citecolor=blue}
\begin{document}

\twocolumn[
  \begin{@twocolumnfalse}
    \maketitle
      
    \begin{abstract}
      \abstractText
      \newline
      \newline
    \end{abstract}
  \end{@twocolumnfalse}
]

\section{Introduction}
\label{intro}

Rotation measurements are utilised for inertial navigation and earth observation exploiting the large enclosed area in fibre-optical and meter-scale ring laser gyroscopes~\cite{Gebauer2020PRL,Schreiber2011PRL,Jekeli2005Nav}.
Atom interferometry offers a different approach for providing absolute measurements of inertial forces with high long-term stability.
Moreover, achieving the necessary areas for competitive performance with matter waves is a long-standing challenge~\cite{Gersemann2020EPJD,Savoie2018SciAdv,Berg2015PRL,Stockton2011PRL,Gauguet2009PRA,Canuel2006PRL,Durfee2006PRL}.

We propose an atom interferometer performing multiple loops in free fall.
Our setup opens the perspective for sensitivities as high as $2\cdot10^{-11}$\,rad/s at 1\,s, comparable to the results of the ring laser gyroscope at the geodetic observatory Wettzell~\cite{Gebauer2020PRL,Schreiber2011PRL}.

The interferometric Sagnac phase shift~\cite{Sagnac1913CRAS} induced by a rotation $\vec{\Omega}$ depends linearly on the area vector $\vec{A}$ as described in the following equation
\begin{equation}
	\Delta\phi_{\mathrm{Sagnac}} = \frac{4\pi E}{\hbar c}\vec{A}\vec{\Omega}
\end{equation}
where $E$ is the energy associated with the atom $E_{\mathrm{at}}=mc^2$ or photon $E_{\mathrm{ph}}=\hbar\omega$, $m$ is the mass of the atom, $\omega$ the angular frequency of the light field, and $c$ the speed of light.
Since $E_{\mathrm{at}}\gg E_{\mathrm{ph}}$, it scales favourably for atoms, motivating early experiments~\cite{Gustavson1997PRL,Riehle1991PRL,Collela1975PRL}, while much larger areas were demonstrated for light~\cite{Schreiber2011PRL}.

In Sagnac sensors based on laser cooled atoms in free fall, light fields driving Raman or Bragg transitions coherently split, deflect, and recombine atomic wave packets in interferometers based on three~\cite{Berg2015PRL,Gauguet2009PRA} or four pulses~\cite{Stockton2011PRL,Canuel2006PRL} with an area of up to $11\,\mathrm{cm}^2$~\cite{Savoie2018SciAdv}.

\begin{figure}[t]
\centering
\includegraphics[width=\columnwidth]{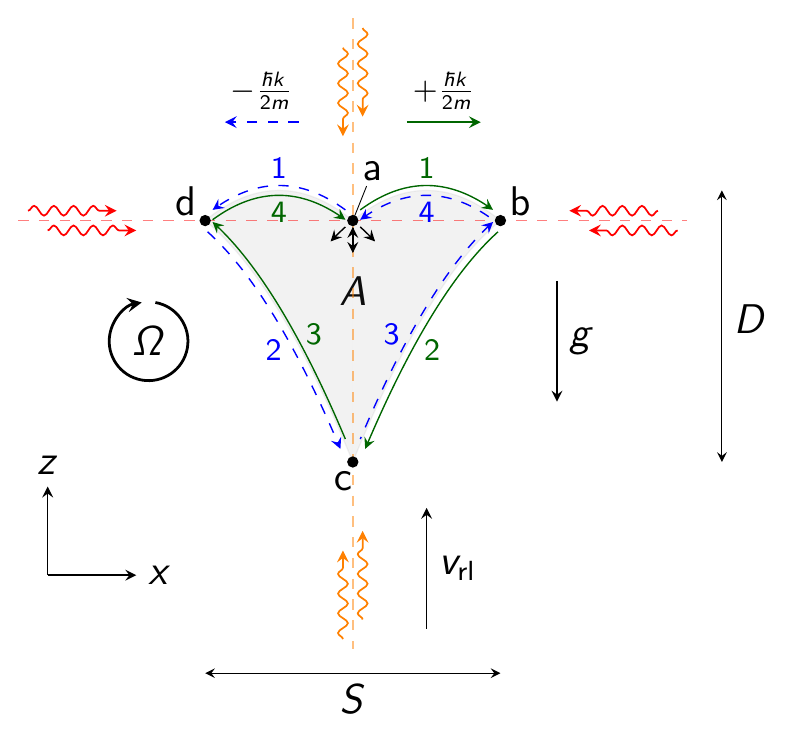}
\caption{Trajectories of the free falling atoms in the interferometer.
Red arrows denote the light fields for splitting, redirecting, and recombination.
Orange arrows indicate the light fields for relaunching the atoms against gravity $g$ with a velocity $v_{\mathrm{rl}}$ enabling operation with a single beam splitting axis (red) and closing the interferometer at its starting point.
The atoms start at (a) where a beam splitting pulse leads to a coherent superposition of two momentum states (blue, green) that separate symmetrically with a recoil velocity of $\pm \hbar k /(2m)$. 
Here, $k$ denotes the effective wave number of the beam splitter (red) and $m$ the atomic mass.
One momentum state follows the green arrows and the second one the dashed blue arrows according to the numbering.
The state deflected in positive x-direction (green) propagates from (a) to (b), (c), (d), and back to (a).
Similarly, but with inverted momentum, the other state (blue) proceeds from (a) to (d), (c), (b), and back to (a), closing one loop.
As a consequence, the interferometer encloses the area $A$ (grey shaded area), rendering it sensitive to rotations $\Omega$. 
Both trajectories meet at (c) where the two momentum states are relaunched at the same time.
Input (up) and output ports (down) of the interferometer are indicated by black arrows below (a).
The maximum wave packet separation is indicated by $S$ and the drop distance by $D$.
}
\label{fig:figure1}
\end{figure}

Another avenue for rotation measurements are wave guides or traps moving atoms in loops, especially for applications requiring compact setups~\cite{Jekeli2005Nav}.
Multiple loops were created in ring traps employing quantum degenerate gases for different purposes including the investigation of superconductive flows~\cite{Ryu2007PRL,Gupta2005PRL}.
Exploiting them for guided atomic Sagnac interferometers in optical or magnetic traps remains an experimental challenge~\cite{Moan2020PRL,Pandey2019Nature,Stevenson2015PRL,Wu2007PRL}.

\begin{figure}[t]
\centering
\includegraphics[width=\columnwidth]{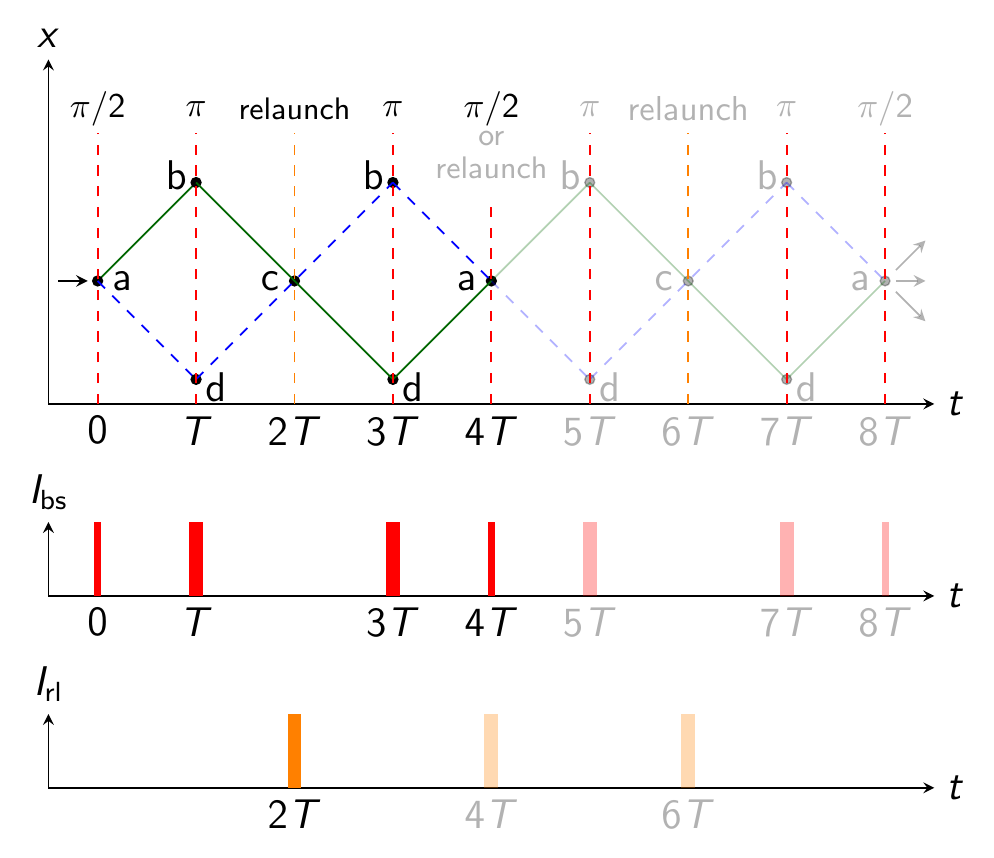}
\caption{Space-time diagram and pulse timings of a multi-loop interferometer.
The upper diagram shows the timing of the $\pi/2$ beam splitting at position (a) in fig.~\ref{fig:figure1}, $\pi$ mirror pulses at position (b) and (d) in fig.~\ref{fig:figure1}, and relaunch at position (c) in fig.~\ref{fig:figure1} as well as the recombination ($\pi/2$) pulse at (a).
Non-opaque lines indicate an implementation with the minimum of two loops ($n=1$) and opaque lines a four-loop interferometer ($n=2$).
The interferometer can be extended to $n$ loops closed with a $\pi/2$ pulse at $4nT$ by introducing relaunches at $r\cdot4T$ for $r\in(1,2,...,n-1)$ (position (a) in fig.~\ref{fig:figure1}).
The lower diagram shows the time-dependent intensities of the beam splitting pulses $I_{\mathrm{bs}}$ and relaunches $I_{\mathrm{rl}}$.
Diagrams are not to scale, neglect pulse shaping and the initial launch before the atoms enter the interferometer.
}
\label{fig:figure2}
\end{figure} 

In our geometry (fig.~\ref{fig:figure1}), atoms are coherently manipulated by two perpendicular light gratings (red and orange dashed lines) to form a multi-loop interferometer~\cite{Schubert2019arXiv}.
Pulsed light fields enable symmetric beam splitting of the atomic wave packets in the horizontal axis (red arrows)~\cite{Gebbe2019arXiv,Ahlers2016PRL,Giese2013PRA,Leveque2009PRL} and relaunching in vertical direction (orange arrows)~\cite{Abend2016PRL}.
This approach offers a variety of advantages: (i) the free-fall time can be tuned to scale the area, (ii) the area is well defined by velocities imprinted during the coherent atom-light interactions, and (iii) the geometry utilises a single axis for beam splitting which avoids the requirement for relative alignment~\cite{Schubert2019arXiv,Savoie2018SciAdv,Tackmann2012NJP}. (iv) It enables multiple loops, (v) and its symmetry suppresses biases due to light shifts associated with the atom-light interaction. (vi) Moreover, our concept in principle allows incorporation of several additional measurements such as local gravity~\cite{Peters1999Nat,Kasevich1991PRL} and tilt of the apparatus~\cite{Ahlers2016PRL} with respect to gravity.
Dual or multi-loop interferometers have also been proposed in the context of terrestrial and space-borne infrasound gravitational wave detection designed for measuring strain rather than rotations~\cite{Canuel2020CQG,El2020EPJQT,Schubert2019arXiv,Graham2016PRD,Hogan2011GRG}.

\section{Multi-loop geometry}
\label{sec:1}

 \begin{table*}[ht]
 \footnotesize
 \renewcommand{\arraystretch}{1.3}
\caption{Comparison of our multi-loop scheme with a four-pulse interferometer and performance estimation.
We assume devices based on rubidium atoms~\cite{Steck2008_Rb87}, a number of $N$ detected atoms, an effective wave number $k$, a pulse separation time $T$ (see fig.~\ref{fig:figure2}), and a contrast $C$, performing $n$ loops.
We denote $A$ as the effectively enclosed area, which scales with $n$.
Both the calculation of $A$ and the sensitivity neglect finite pulse durations.
The maximum trajectory separation is given by $S$.
For the estimation of the drop distance $D$ with regard to a compact scenario we allow for additional time of 6\,ms for the momentum transfer.
Our interferometer cycle time is denoted by $t_{\mathrm{c}}$ and we state the sensitivity in the shot-noise limit according to eq.~\ref{eq:noise}.
In the first four rows we compare multi-loop interferometers to four-pulse geometries without a relaunch~\cite{Savoie2018SciAdv,Stockton2011PRL,Canuel2006PRL}, emphasising differences in the parameters in blue, and neglecting atom losses and contrast reduction due to imperfect beam splitters.
For the lower two rows, we assume a simple model in which the contrast for multiple loops $C(n)$ depends on the contrast for a single loop $C(1)$ and scales as $C(n)=C(1)^{n}$.
Furthermore, our model reduces the number of detected atoms by a factor $l^{n-1}$ with $l=0.9$ for additional loops to take inefficiencies in the atom-light interactions into account.
}
\centering
\begin{tabular}{p{2.5cm} | p{1cm} p{1.1cm} p{0.5cm} p{0.2cm} p{0.5cm} p{1.2cm} p{0.8cm} p{1.4cm} p{1.6cm} p{1.5cm} }
     sensor features & $N$ & $k$ & $T$ & $n$ & $C$ & $A$ & $t_{\mathrm{c}}$ & $S$ & $D$ & sensitivity \\
 &  & $\left[ \frac{2\pi}{780\,\mathrm{nm}} \right]$ & [ms] &  &  & $\left[ \mathrm{m}^2 \right]$ & [s] & [m] & [m] & $\left[ \frac{\mathrm{rad/s}}{\sqrt{\mathrm{Hz}}} \right]$ \\
\midrule 
 
   1: multi loop & $10^5$ & \textcolor{blue}{40} & 10 & 10 & 1 & $4.6\cdot10^{-5}$ & 1.6 & \textcolor{blue}{$2.4\cdot10^{-3}$} & \textcolor{blue}{$2.8\cdot10^{-3}$} & $3.2\cdot10^{-8}$ \\
   1: four pulse & $10^5$ & \textcolor{blue}{350} & 10 & - & 1 & $4\cdot10^{-5}$ & 1.24 & \textcolor{blue}{$2.1\cdot10^{-2}$} & \textcolor{blue}{$5\cdot10^{-3}$} & $3.2\cdot10^{-8}$ \\
   2: multi loop & $4\cdot10^5$ & 20 & 250 & 10 & 1 & \textcolor{blue}{$3.6\cdot10^{-1}$} & \textcolor{blue}{11.8} & $3\cdot10^{-2}$ & 0.7 & \textcolor{blue}{$5.5\cdot10^{-12}$} \\
   2: four pulse & $4\cdot10^5$ & 28 & 189 & - & 1 & \textcolor{blue}{$2.1\cdot10^{-2}$} & \textcolor{blue}{2.8} & $3.1\cdot10^{-2}$ & $0.7$ & \textcolor{blue}{$4.2\cdot10^{-11}$} \\

 \midrule
   compact & $5.9\cdot10^4$ & 40 & 10 & 6& 0.53 & $2.8\cdot10^{-5}$ &  1.44 & $2.4\cdot10^{-3}$ & $2.8\cdot10^{-3}$ & $1.2\cdot10^{-7}$ \\
   high sensitivity& $2.9\cdot10^5$ & 20 & 250 & 4 & 0.66 & $1.4\cdot10^{-1}$ & 5.8 & $3\cdot10^{-2}$ & 0.7 & $1.7\cdot10^{-11}$ \\ 
  \bottomrule                             
\end{tabular}
\label{tab:tab1}
\end{table*}

The trajectories in our interferometer are detailed in fig.~\ref{fig:figure1}.
Initially, an atomic wave packet is launched vertically.
On its upward way, the wave packet interacts with the horizontal beam splitter with an effective wave number $k$ forming two wave packets drifting apart with a momentum of $\pm\hbar k/(2m)$.
After a time $T$, the horizontally oriented light field (red) inverts the movement of the atoms on its axis.
On their way down due to gravity, the vertically oriented light field (orange) relaunches the atoms~\cite{Abend2016PRL} at the lowest point of the interferometer at $2T$ reversing their momentum to move upward.
The atoms pass the horizontal atom-light interaction zone again at $3T$ where they are deflected towards each other and cross falling downwards at $4T$ completing their first loop.
In order to start the next loop, they are relaunched (see fig.~\ref{fig:figure2}).
Repetition of the procedure determines the number of loops $n$.
After the last loop, the interferometer is closed by flashing a beam splitter pulse instead of an upward acceleration. 

The resulting area determining the sensitivity to rotations depends on the total time of the interferometer $4T$, the effective wave vector $\vec{k}$, local gravity $\vec{g}$, and leads to the Sagnac phase
\begin{equation}
	\Delta\phi_{\mathrm{Sagnac}} = n\cdot4(\vec{k}\times\vec{g})\vec{\Omega} T^3 \label{eq:phidlai}
\end{equation}
calculated with the methods outlined in ref.~\cite{Hogan2008arXiv,Bongs2006APB,Borde2004GRG} and similar as in refs.~\cite{Savoie2018SciAdv,Stockton2011PRL,Canuel2006PRL,Dubetsky2006PRA}.
The relaunch velocity $v_{\mathrm{rl}}=|\vec{v}_{\mathrm{rl}}|=3gT$ with $g=|\vec{g}|$ is aligned parallel to gravity and is chosen to close the atom interferometer at its starting point. 
In this configuration, the area is given by
\begin{equation}
	A = n\cdot2\frac{\hbar k}{m}g T^3. \label{eq:A}
\end{equation}
It can be enlarged by a higher transverse momentum $\hbar k=\hbar|\vec{k}|$, e.g. by transferring more photon recoils, and by increasing the free fall time $4T$ of the interferometer.

Enlarging the number of loops by a factor $n$ effectively increases the enclosed area without changing the dimensions of the geometry, defined by the maximum wave packet separation in the horizontal axis $S=\hbar k T/m$ and the drop distance in the vertical axis $D=(3T/2)^2\cdot g/2$.
The relaunch at $4T$ (or multiples of $4T$ for more than four loops, position (a) in fig.~\ref{fig:figure1}) reuses the same light field (orange arrows) as for the first relaunch at $2T$ (position (c) in fig.~\ref{fig:figure1}) and does thus not add complexity.

Typically, the cycle of an atom interferometer consists of the generation and preparation of the atomic ensembles during the time $t_{\mathrm{prep}}$, the interferometer time which for our geometry reads $n\cdot 4T$, and detecting the population of the output ports within the time $t_{\mathrm{det}}$.
This leads to a total cycle time of $t_c=t_{\mathrm{prep}}+n\cdot 4T+t_{\mathrm{det}}$.
Using the phase shift given in eq.~\ref{eq:phidlai}, the shot-noise limited sensitivity for rotations $\Omega_{y}$ with $N$ atoms and a contrast $C$ is given by
\begin{equation}
	\sigma_{\Omega}(t) = \frac{1}{C\sqrt{N}\cdot n\cdot(4kgT^3)}\sqrt{\frac{t_{\mathrm{prep}}+n\cdot 4T+t_{\mathrm{det}}}{t}} \label{eq:noise}
\end{equation}
for an averaging time $t$ corresponding to multiples of the cycle time $t_{\mathrm{c}}$.
Consequently, an interferometer with a small free fall time $4T\ll t_{\mathrm{prep}}+t_{\mathrm{det}}$ benefits more from multiple loops with a scaling of $\sim 1/n$ in eq.~\ref{eq:noise} than other scenarios with $4T\thickapprox t_{\mathrm{prep}}+t_{\mathrm{det}}$ that scale as $\sim 1/\sqrt{n}$.
Implementing an interferometer time $n\cdot 4T > t_{\mathrm{prep}}$ can enable a continuous scheme by sharing $\pi/2$ pulse between subsequent interferometers~\cite{Savoie2018SciAdv,Dutta2016PRL}.
Here, our geometry offers the possibility to compensate smaller $T$ with an appropriate $n$. 

\section{Spurious phase shifts}
\label{sec:Spurious phase shifts}

In multi-loop interferometers, the sensitivity to DC accelerations and phase errors depending on the initial position and velocity is suppressed~\cite{Hogan2011GRG,Dubetsky2006PRA}.
However, a non-ideal pointing of the relaunch velocity $\vec{v}_{\mathrm{rl}}$ may introduce spurious phase shifts in a real setup.
We consider small deviations $\alpha=|\vec{v}_{\mathrm{rl}}\times\vec{e}_{x}|/(|\vec{v}_{\mathrm{rl}}||\vec{e}_{x}|)$ and $\beta=|\vec{v}_{\mathrm{rl}}\times\vec{e}_{y}|/(|\vec{v}_{\mathrm{rl}}||\vec{e}_{y}|)$ in a double-loop configuration.
Here, $\vec{e}_{x}=\vec{k}/|\vec{k}|$ denotes the unit vector in x-direction and $\vec{e}_{y}=(\vec{k}\times\vec{g})/(|\vec{k}||\vec{g}|)$ denotes the unit vector in y-direction (see fig.~\ref{fig:figure1}).

If the timing of the relaunch is not ideally centred around $2T$, but shifted by $\delta\tau$, coupling to non-zero $\alpha$ leads to the phase shift~\cite{Schubert2019arXiv}
\begin{equation}
	\Delta\phi_{\alpha,\tau}=-kv_{\mathrm{rl}}\alpha\delta\tau =-3kgT\alpha\delta\tau. \label{eq:tau}
\end{equation}
Provided the pointing of the relaunch velocity is adjustable (e.g. with a tip-tilt mirror controlling the alignment of the light field for relaunching) $\alpha$ and $\delta\tau$ can be adjusted by iteratively scanning both.

In addition, tilting the relaunch vector induces phase shifts resembling those of a three-pulse or Mach-Zehnder-like interferometer~\cite{Hogan2008arXiv,Bongs2006APB} by coupling to gravity gradients $\Gamma$ and rotations $\vec{\Omega}$.
These contributions read
\begin{equation}
	\Delta\phi_{\alpha,\Gamma} = \vec{k}\Gamma \vec{v}_{\mathrm{rl}}T^3 = 3k\alpha\Gamma_{x} gT^4, \label{eq:gamma}
\end{equation}
with $\Gamma_x=\vec{e}_{x}\Gamma\vec{e}_{x}$ and
\begin{equation}
	\Delta\phi_{\beta,\Omega} = 2 \left( \vec{k}\times \vec{v}_{\mathrm{rl}} \right)\cdot\vec{\Omega}T^2 = 6k\beta g\Omega_{z}T^3, \label{eq:omega}
\end{equation}
corresponding to a spurious sensitivity to a rotation $\Omega_{z}=\vec{e}_z\cdot\vec{\Omega}$.

Scanning the interferometer time $4T$ enables an iterative procedure to minimise spurious phase shifts by optimising the contrast~\cite{Tackmann2012NJP}.

 \begin{table}[t]
 \footnotesize
\caption{Requirements on the pointing of the relaunch.
The angles $\alpha_{\delta\tau}$, $\alpha_{\Gamma}$ and $\beta$ are calculated to induce contributions (eqs.~\ref{eq:tau}, \ref{eq:gamma}, and~\ref{eq:omega}) by a factor of $10\cdot{n}$ below the shot-noise limit for the scenarios in the lower rows of tab.~\ref{tab:tab1} at 1\,s.
We assume $\delta\tau=10\,\mathrm{ns}$ for a typical experiment control system~\cite{Asenbaum2020PRL}, Earth's gravity gradient $\Gamma_{x}=1.5\cdot10^{-6}\,\mathrm{s}^{-2}$, and Earth's rotation rate $\Omega_{z}=7.27\cdot10^{-5}\,\mathrm{rad/s}$~\cite{Hogan2008arXiv,Bongs2006APB}.
Values are given in $\,\mathrm{rad}$.
}
\centering
\begin{tabular}{p{2.2cm} | p{1.4cm} p{1.5cm} p{1.4cm} }
      & $\alpha_{\delta\tau}$ & $\alpha_{\Gamma}$ & $\beta$ \\
  \midrule
   compact & $1.3\,\cdot10^{-4}$ & $<0.1$ & $9.4\,\cdot10^{-5}$ \\
   high-sensitivity& $6\,\cdot10^{-6}$ & $2.5\,\cdot10^{-6}$ & $6.6\,\cdot10^{-9}$ \\
\bottomrule                             
\end{tabular}
\label{tab:tab2}
\end{table}

\section{Perspectives of the performance}
\label{sec:Perspectives of the performance}

Apart from the (re-)launch, our scheme with $n=1$ resembles four-pulse interferometers~\cite{Savoie2018SciAdv,Stockton2011PRL,Canuel2006PRL,Dubetsky2006PRA} in geometry and scale factor (eqs.~\ref{eq:phidlai} and~\ref{eq:A}).
Hence, we show the advantages of our method with respect to a four-pulse interferometer for two design choices.
For both interferometers in tab.~\ref{tab:tab1} (upper four rows), we assume ideal contrast, no losses of atoms, as well as shot-noise limited sensitivities (see eq.~\ref{eq:noise}).
(1) Matching the free-fall time $T$ and sensitivity for the multi-loop and four-pulse sensor, the four-pulse interferometer requires a larger photon momentum transfer in horizontal direction.
Consequently, the size $S\cdot{D}$ of
the multi-loop geometry is by a factor of 15 smaller (emphasised in blue in tab.~\ref{tab:tab1}).
(2) Aiming for similar dimensions $S$ and $D$, we obtain a nearly an order of magnitude higher sensitivity for our multi-loop geometry at the cost of an increased cycle time.

Showcasing more realistic scenarios, we consider a simple model for losses of atoms and reduction of contrast in dependence on the number of loops as summarised in tab.~\ref{tab:tab1} (lower rows).
The latter can result from inhomogeneities of the light fields~\cite{Kim2020arXiv,Gebbe2019arXiv,Bade2018PRL,Zhang2016PRA,Parker2016PRA,Estey2015PRL,Kovachy2015Nature}.
According to our estimations, our method would still lead to a compact sensor with a sensitivity of $1.2\cdot10^{-7}\,(\mathrm{rad/s})/\sqrt{\mathrm{Hz}}$ within a volume of 20\,mm$^3$ for the interferometer, and a highly sensitive, but larger device with $1.7\cdot10^{-11}\,(\mathrm{rad/s})/\sqrt{\mathrm{Hz}}$ within a meter-sized vacuum vessel, comparable to the performance of large ring laser gyroscopes~\cite{Schreiber2011PRL}.

Multiple experiments investigated beam splitting as well as relaunch operations as required for our scheme.
They realised the transfer of large momenta with subsequent pulses or higher order transitions~\cite{Rudolph2020PRL,Parker2018Science,Kovachy2015Nature,Chiow2011PRL,Mueller2008PRL,McGuirk2000PRL} and their combination with Bloch oscillations~\cite{McDonald2014EPL,Muller2009PRL,Clade2009PRL}.
The implementation of symmetric splitting~\cite{Pagel2019arXiv,Gebbe2019arXiv,Plotkin2018PRL,Ahlers2016PRL,Leveque2009PRL} was demonstrated with an effective wave number corresponding to $408$ photon recoils in a twin-lattice atom interferometer~\cite{Gebbe2019arXiv}.
A similar procedure enabled the relaunch of atoms~\cite{Abend2016PRL}.
The requirement for high efficiency implies using atomic ensembles with very low residual expansion rates~\cite{Szigeti2012NJP} as enabled by delta-kick collimation of evaporated atoms~\cite{Asenbaum2020PRL,Kovachy2015PRL} and Bose-Einstein condensates~\cite{Gebbe2019arXiv,Abend2016PRL,Rudolph2016Diss,Muntinga2013PRL,Dickerson2013PRL}.
In addition, interferometers exploiting such ensembles may benefit from the suppression systematic of uncertainties~\cite{Heine2020EPJD,Karcher2018NJP,Louchet2011NJP}.
Fountain geometries utilised launch techniques compatible with these ensembles~\cite{Asenbaum2020PRL,Abend2016PRL,Kovachy2015PRL,Dickerson2013PRL}.
Rapid generation of Bose-Einstein condensates with $10^5$ atoms was demonstrated~\cite{Roy2016PRA,Stellmer2013PRA} and realised with atom chips in 1\,s~\cite{Becker2018Nature,Rudolph2015NJP} which we adopted for our estimation.

Reaching the shot-noise limited sensitivity implies a restriction on tilt instability as detailed in tab.~\ref{tab:tab2} due to couplings in eqs.~\ref{eq:tau}, \ref{eq:gamma}, and \ref{eq:omega}.
It is met at the modest level of $0.1\,\mathrm{mrad/}\sqrt{\mathrm{Hz}}$ for the compact scenario and at $7\,\mathrm{nrad/}\sqrt{\mathrm{Hz}}$ for high-sensitivity~\footnote{Dedicated vibration isolation systems demonstrated a noise floor of $1\,\mathrm{nrad/}\sqrt{\mathrm{Hz}}$ in a frequency range of 1\,Hz to 100\,Hz~\cite{Bergmann2018Diss}. Alternatively, and similar as in a large ring laser gyroscope~\cite{Gebauer2020PRL,Schreiber2011PRL}, tiltmeters with a resolution of sub\,nrad~\cite{lgm2020} may enable post correction methods.}.

\section{Conclusion and discussion}
\label{sec:Conclusion}

We presented our concept for an atomic gyroscope capable of performing multiple loops by exploiting light pulses for beam splitting and relaunching atoms with the perspective of reaching unprecedented sensitivities for rotations. 
It offers unique scalability in a limited size of a sensor head.
Key elements as the symmetric beam splitting~\cite{Gebbe2019arXiv,Ahlers2016PRL}, relaunch~\cite{Abend2016PRL}, as well preparation of the ultracold atoms~\cite{Becker2018Nature,Hardman2016PRL,Kovachy2015PRL,Rudolph2015NJP,Muntinga2013PRL} have already been demonstrated.
The tools for coherent manipulation in our scheme additionally allow for the implementation of geometries for a tiltmeter~\cite{Ahlers2016PRL} and a gravimeter~\cite{Abend2016PRL,Peters1999Nat,Kasevich1991PRL}.
We showed the perspective for compact setups, which can be scaled up to compete with large ring laser gyroscopes~\cite{Gebauer2020PRL,Schreiber2011PRL}.
This might enable the detection of multiple rotational components in a single set-up by adding a second orthogonal beam splitting axis, and sensitivities as required for measuring the Lense-Thirring effect~\cite{Everitt2011PRL,Jentsch2004GRG,Schiff1960PRL,Lense1918PZ}.

\section*{Acknowledgements}
The presented work is supported by the CRC 1227 DQmat within the projects B07 and B09, the CRC 1464 TerraQ within the projects A01, A02 and A03, the German Space Agency (DLR) with funds provided by the Federal Ministry of Economic Affairs and Energy (BMWi) due to an enactment of the German Bundestag under Grant No. DLR 50WM1952 (QUANTUS-V-Fallturm), 50WP1700 (BECCAL), 50RK1957 (QGYRO), and the Verein Deutscher Ingenieure (VDI) with funds provided by the Federal Ministry of Education and Research (BMBF) under Grant No. VDI 13N14838 (TAIOL).
Funded by the Deutsche Forschungsgemeinschaft (DFG, German Research Foundation) under Germany’s Excellence Strategy – EXC-2123 QuantumFrontiers – 390837967.
D.S. acknowledges support by the Federal Ministry of Education and Research (BMBF) through the funding program Photonics Research Germany under contract number 13N14875.
We acknowledge financial support from “Nieders{\"a}chsisches Vorab” through “F{\"o}rderung von Wissenschaft und Technik in Forschung und Lehre“ for the initial funding of research in the new DLR-SI Institute and through the ``Quantum- and Nano-Metrology (QUANOMET)'' initiative within the project QT3.

\bibliographystyle{unsrt}
\begin{footnotesize}

\end{footnotesize}
\end{document}